\pdfoutput=1
\documentclass[cernpreprint, UKenglish, texlive=2016, texmf]{atlasdoc}
\usepackage[subcaption]{atlaspackage}
\usepackage{atlasbiblatex}
\usepackage{atlasphysics}

\usepackage{multirow}
\usepackage{longtable}
\usepackage{booktabs}
\usepackage{arydshln}

\usepackage{paper-defs}

\AtlasTitle{Measurement of differential cross sections of isolated-photon plus heavy-flavour jet production in \pp collisions at \rts = 8~\TeV\ using the ATLAS detector}

\AtlasAbstract{
This Letter presents the measurement of differential cross sections of isolated prompt photons produced in association with a $b$-jet or a $c$-jet. These final states provide sensitivity to the heavy-flavour content of the proton and aspects related to the modelling of heavy-flavour quarks in perturbative QCD. The measurement uses proton--proton collision data at a centre-of-mass energy of 8~\TeV\ recorded by the ATLAS detector at the LHC in 2012 corresponding to an integrated luminosity of up to 20.2~\ifb. The differential cross sections are measured for each jet flavour with respect to the transverse energy of the leading photon in two photon pseudorapidity regions: $|\etag|<1.37$ and $1.56<|\etag|<2.37$. The measurement covers photon transverse energies $25<\etg<400$~\GeV\ and $25<\etg<350$~\GeV\ respectively for the two $|\etag|$ regions. For each jet flavour, the ratio of the cross sections in the two $|\etag|$ regions is also measured. The measurement is corrected for detector effects and compared to leading-order and next-to-leading-order perturbative QCD calculations, based on various treatments and assumptions about the heavy-flavour content of the proton. Overall, the predictions agree well with the measurement, but some deviations are observed at high photon transverse energies. The total uncertainty in the measurement ranges between 13\% and 66\%, while the central $\gamma+b$ measurement exhibits the smallest uncertainty, ranging from 13\% to 27\%, which is comparable to the precision of the theoretical predictions.
}

\author{The ATLAS Collaboration}

\AtlasRefCode{STDM-2017-02}

\PreprintIdNumber{CERN-EP-2017-217}

\AtlasJournalRef{Phys.\ Lett.\ B 776 (2018) 295}

\AtlasDOI{10.1016/j.physletb.2017.11.054}

\hypersetup{pdftitle={ATLAS document},pdfauthor={The ATLAS Collaboration}}

\begin{document}

\maketitle

\clearpage

\section{Introduction}
\label{sec:Intro}

The production of isolated prompt photons in association with a jet containing a $b$- or $c$-hadron provides a testing ground for perturbative quantum chromodynamics (pQCD), the content of the proton and the treatment of heavy quarks in matrix element (ME) and parton shower (PS) computations. Prompt photons, which refer to those not arising from hadron decays, are targeted by requiring that their signals are isolated, i.e.\ well separated from other energetic signals.
The most recent measurements of these final states were performed at the Tevatron proton--antiproton collider by the D0 \cite{D0:2012ea,D0:2012gw} and CDF \cite{CDF:2013ama} collaborations. The Large Hadron Collider (LHC) produces proton--proton (\pp) collisions at much higher centre-of-mass energies. Compared to the proton--antiproton collisions of the Tevatron, these collisions exhibit smaller contributions from $t$-channel quark--antiquark processes, allowing other processes sensitive to the heavy-quark content of the proton to play a more significant role.\footnote{In the context of a photon+jet final state, $s$-channel quark--antiquark processes are suppressed due to the isolation requirement imposed on the photon.}

Prompt photons ($\gamma$) can be used as a colourless non-hadronizing probe of parton dynamics that yields a clean experimental signature~\cite{STDM-2011-28,diphoton7TeV2,yJet7TeV,incphoton7TeV,incphoton8TeV,yJet8TeV,incphoton13TeV,diphoton8TeV,CMS6,CMS4,CMS3,CMS2,CMS1}. Processes containing final state $b$-  or $c$-quarks play an important role in many LHC physics analyses and therefore the accuracy of the description of this heavy-flavour (HF) content of the proton must be investigated \cite{ScalarCharmTheory,TopPhysStatus,HiggsBranching,ScalarCharm,SUSY,}. HF jets are defined as jets which contain either a $b$- or $c$-hadron. 

At the LHC, prompt photons arise mainly through the Compton process, initiated by a quark ($q$) and a gluon ($g$), $qg \rightarrow q\gamma$. HF quarks arise in the proton through either extrinsic or intrinsic mechanisms. Extrinsic refers to HF quarks arising through perturbative mechanisms in the proton, while intrinsic refers to non-perturbative mechanisms. Presently, global parton distribution function (PDF) fits show that HF quarks in the proton are almost entirely extrinsic, however 
non-zero values of the intrinsic contribution have not been ruled out \cite{Gennady}.
The photon transverse energy observable provides sensitivity to these effects, by taking advantage of its precise calibration while integrating over the less precise jet kinematic observables. The effect of intrinsic HF quarks in the PDF would be manifest at large Bjorken-$x$, which in the context of this measurement would give rise to larger cross-section values at large absolute photon pseudorapidities, $|\etag|$, with high photon transverse energy, $\etg$.\footnote{ATLAS uses a right-handed coordinate system with its origin at the nominal interaction point (IP) in the centre of the detector and the $z$-axis along the beam pipe. The $x$-axis points from the IP to the centre of the LHC ring and the $y$-axis points upward. Cylindrical coordinates $(r,\phi)$ are used in the transverse plane, $\phi$ being the azimuthal angle around the $z$-axis. The pseudorapidity is defined in terms of the polar angle $\theta$ as $\eta=-\ln\tan(\theta/2)$. The rapidity is defined as $y=(1/2) \cdot \ln {( (E+p_{z})/(E-p_{z}) )}$.} Due to their smaller mass, $c$-quarks are more sensitive to these effects than $b$-quarks.

As the value of the mass of the $b$-quark, $m_b$, is much greater than the non-perturbative scale of QCD, it can be included explicitly in pQCD calculations. The calculations of the \yb cross sections can thus be done in two different schemes: the four-flavour scheme (4F) and the five-flavour scheme (5F) \cite{Scheme4F5F}. 
In the 4F scheme, the $u$-, $d$-, $s$- and $c$-quarks are treated as massless in the ME, while the $b$-quark is treated as massive. PDFs describing the proton with only the lightest four quarks are used in the calculations. Most of the $b$-quarks are dynamically generated in the matrix element through the splitting of a gluon. 
In the 5F scheme, the $b$-quark is also treated as massless and PDFs describing the proton with these five quarks are used in the calculations. The $b$-quark can thus be an initial-state quark in the matrix element. As such, the diagrams considered at a particular order in pQCD differ between the 4F and 5F schemes. In the 4F scheme, the mass of the $b$-quark gives rise to terms proportional to $\log({Q^2}/{m_b^2})$. At energies far above $m_b$, these logarithmic terms are large and spoil the convergence of the perturbative series. The 5F scheme avoids this issue since these logarithmic terms are resummed into the $b$-quark PDF. As such the 5F scheme is expected to give a better description of processes at energies far above $m_b$. In a calculation to all orders in pQCD, however, both schemes should yield the same result.

Theoretical predictions using leading-order (LO) and next-to-leading-order (NLO) pQCD matrix element calculations, interfaced to a PS, are computed using different PDF sets and compared to the measurement. In addition to the nominal sets with no intrinsic charm component in the 5F scheme, predictions are made using sets in the 4F scheme for \yb and sets that incorporate various degrees of intrinsic charm contribution for \yc. 

This Letter presents a fiducial differential measurement of the production cross section of a prompt isolated photon in association with a $b$-jet or a $c$-jet in $pp$ collisions using the ATLAS detector. The transverse photon energy, $\etg$, is required to satisfy $\etg > 25$~\GeV, the jet transverse momentum, $\ptj$, is required to satisfy $\ptj > 20$~\GeV\ and the absolute pseudorapidity of the jet, $|\etaj|$, is required to satisfy $|\etaj| < 2.5$. In each bin of the measurement a background-enriched sideband technique is used to extract the prompt photon signal, while a template fit of a neural-network jet flavour-tagging discriminant is used to extract the HF signal. The measured signal is then corrected for detector effects, mapping it from the detector level to the particle level. The measurement is performed in bins of $\etg$ for two regions of $|\etag|$: the central region with $|\etag| < 1.37$ reaching up to $400$~\GeV\ in $\etg$ and the forward region with $1.56 < |\etag| < 2.37$ reaching up to $350$~\GeV\ in \etg. The ratios of the cross sections in the central to the forward regions are also presented for each flavour, as systematic and theoretical uncertainties that are correlated between the two regions then cancel out.

\section{ATLAS detector}
\label{sec:Detector}
The ATLAS detector \cite{ATLAS} is designed to measure the particles produced by the collisions provided by the LHC with almost complete solid angle coverage of the collision point. 
The inner detector (ID), immersed in a 2 T axial magnetic field provided by an encompassing thin superconducting solenoid, is located nearest to the beam pipe and comprises a high-granularity pixel detector, a silicon microstrip tracker and a straw-tube transition radiation tracker. The ID provides tracking and vertexing information, which plays a crucial role in this measurement identifying photons and HF decay vertices associated with jets. Its acceptance, up to $|\eta| = 2.5$, imposes the upper bound on the $|\eta|$ acceptance of the analysis. 
The electromagnetic calorimeter (ECAL) surrounds the ID and is used to measure electromagnetic showers. Within the $|\eta|$ acceptance of this analysis, the ECAL is a high-granularity lead/liquid-argon sampling calorimeter using an accordion geometry that provides complete azimuthal coverage and comprises three radial layers augmented by a thin presampler. 
The presampler, which covers $|\eta| < 1.8$ and is located in front of the ECAL strips, is used to measure early electromagnetic showers. 
The innermost ECAL layer is the thinnest and uses highly segmented strips in $\eta$, which help to characterize shower shapes. The second layer is the thickest with a coarser granularity and collects most of the photon energy. The third layer is the least granular and is used to correct high-energy signals for leakage. Between the ECAL barrel and endcap detectors there is a transition region, located at $1.37 < |\eta| < 1.56$, where the photon reconstruction and identification are poorer. 
The hadronic calorimeter (HCAL) encloses the ECAL and is used to measure hadronic showers. The HCAL consists of a steel/plastic-scintillator sampling calorimeter for $|\eta| < 1.7$ and a copper/liquid-argon sampling calorimeter for $1.5 < |\eta| < 3.2$. 
Surrounding the HCAL is the muon spectrometer, equipped with large superconducting air-core toroidal magnets and comprising separate sets of detectors for triggering and for precision muon track reconstruction.
A three-level trigger system is used to select photon signals. The first-level trigger is a coarse-granularity hardware-based trigger that limits the event rate to 75 kHz. The second- and third-level triggers are software-based and make use of the full detector granularity, reducing the event rate to about 400 Hz.

\section{Monte Carlo simulations and theoretical predictions}
\label{sec:Theory}

The \SHERPA~1.4.5~\cite{Gleisberg:2008ta} and \PYTHIA~8.160~\cite{Sjostrand:2007gs} Monte Carlo (MC) event generators are used to simulate signal events of prompt photons accompanied by jets at LO in pQCD. The cross sections predicted by both these MC generators are compared to the measured values. \SHERPA is also used to derive correction factors used in the data analysis while \PYTHIA is used to assess some modelling uncertainties.

\SHERPA is used in the ME+PS prescription~\cite{Hoeche:2009rj} to generate events containing a photon and a parton, with up to three additional partons. All photon emissions are effectively simulated by the combination of the tree-level matrix elements including additional partons and the parton shower~\cite{Siegert:2016bre}. Collinear divergences from the photon are regularized by requiring a minimum angular separation of $\Delta R = \sqrt{(\Delta \eta)^2 + (\Delta \phi)^2 } = 0.3$ between the photon and any parton. \PYTHIA is used to generate $2\to2$ events containing either a photon and a parton or two partons, where photons in the latter case are produced in the initial- and final-state radiation. The non-perturbative QCD models used for the parton shower, the hadronization and the hadron decays are different between \SHERPA~\cite{Schumann:2007mg,Winter:2003tt} and \PYTHIA~\cite{Sjostrand:2004ef,Andersson:1983ia}. Both generators also include the effects of the underlying event. For the event generation with \SHERPA, the five-flavour CT10 PDF set~\cite{Lai:2010vv} is used in conjunction with dedicated parton shower tuning developed by the \SHERPA authors. For the event generation with \PYTHIA, the five-flavour LO CTEQ6L1 PDF set~\cite{Pumplin:2002vw} is used with the AU2 set of tuned MC parameters~\cite{ATLAS:2012uec}. In the calculation of the matrix element, \SHERPA uses massive quarks, thus its calculations are in a massive 5F scheme, while \PYTHIA uses massless quarks, thus its calculations are in a standard 5F scheme. Both generators use massive $b$- and $c$-quarks in the parton showers. A \textsc{Geant4} simulation~\cite{Agostinelli:2002hh} of the ATLAS detector~\cite{Aad:2010ah} is used to simulate the interactions between the particles and the detector. During the simulation, the signal events are overlaid with multiple $pp$ collisions generated with the soft QCD processes of \PYTHIA using the A2 set of tuned MC parameters~\cite{ATLAS:2012uec} and the MSTW2008LO PDF set~\cite{Martin:2009iq}. The resulting events are scaled to the integrated luminosity measured in the data. They are also weighted to reproduce the observed distribution in data of the number of reconstructed primary vertices and the size of the luminous region along the beam axis.

In addition, NLO pQCD particle-level predictions calculated using \MGMCatNLO~2.3.3~\cite{Alwall:2014hca} in the 5F scheme, interfaced to \PYTHIA~8.212~\cite{Sjostrand:2014zea} in the NLO+PS prescription, are used to interpret the measurement. For the \yb cross section, the 4F scheme is also considered. In the 5F scheme, events containing a photon and a jet are generated. After hadronization, only events with a jet containing a HF hadron are considered. In the 4F scheme, $b$-quarks are pair-produced from a gluon, hence events containing a photon and two $b$-quarks are generated. In contrast to the 5F scheme, all generated events are considered as they all contain at least one $b$-quark at parton level. In both schemes, the photon collinear divergences are regularized in the matrix element by requiring the photons to pass a Frixione isolation cut~\cite{Frixione:1998jh}: $E_\text{T}^\text{iso}(\delta)<\epsilon\etg\left({(1-\cos\delta)}/{(1-\cos\delta_0)}\right)^n$ with parameters $\delta_0=0.4$, $n=1$ and $\epsilon=1$, where $E_\text{T}^\text{iso}(\delta)$ is the sum of the transverse energies of the particles around the photon up to an angular separation of $\delta$ in the $\eta$--$\phi$ space. The renormalization and factorization scales, \muR and \muF respectively, are chosen to be equal to the transverse mass of the clustered jets, obtained after all final-state particles from the matrix element are $k_t$-clustered~\cite{Alwall:2007fs} into jets. This choice follows the recommendations in Ref.~\cite{Alwall:2014hca} when interfacing the \MGMCatNLO calculations to \PYTHIA. The \yb predictions use the NNPDF3.0nlo\footnote{NNPDF3.1nlo PDF sets with a 4F description of the proton were not available when this analysis was conducted. The PDF sets with a 5F description of the proton were found to produce consistent results between NNPDF3.0nlo and NNPDF3.1nlo.} 4F and 5F PDF sets~\cite{Ball:2014uwa}, while the \yc predictions use NNPDF3.1nlo~\cite{Ball:2017nwa} and CT14nnlo~\cite{Dulat:2015mca}. The NNPDF3.1nlo sets include a set with a charm contribution fitted to data in the global PDF fit, equivalent to intrinsic charm contributing 0.26\% of the proton momentum, and another with only perturbative charm. CT14nnlo provides two sets using the BHPS model~\cite{IC} that include intrinsic charm contributions~\cite{Hou:2017khm}: one with 0.6\% of the proton momentum assigned to intrinsic charm, BHPS1, and one with 2.1\%, BHPS2. The PDF sets include the running of the strong coupling constant, using a value at the energy scale of the mass of the $Z$ boson of ${\alphas(M_{Z})}=0.118$. This treatment of \alphas is used in both the PDFs and the matrix elements. The electromagnetic coupling constant is set to $\alpha$ = 1/137 and its running is not included in the calculations~\cite{Czarnecki:1998tn}.

Three types of uncertainties are considered in the NLO predictions. The scale uncertainty is assessed by multiplying or dividing by a factor of two \muR and \muF, separately and simultaneously. The envelope of the deviations from the nominal prediction is taken as the uncertainty. The uncertainty in the PDF sets is propagated to the cross sections using the prescribed eigenvector reduction approach for the CT14nnlo sets, which gives an uncertainty at the 90\% confidence level. For the NNPDF3.0nlo and NNPDF3.1nlo PDF sets, the PDF uncertainty is assessed through the use of PDF replicas. The uncertainty due to the \alphas value used in the predictions is assessed by varying up or down its value at the energy scale of the mass of the $Z$ boson by 0.002 simultaneously in the matrix element and the PDF sets, resulting in an uncertainty in the cross sections at the 90\% confidence level. In all cases, the uncertainties are reported at the 68\% confidence level in the comparisons to data. The total theoretical uncertainty in the NLO predictions is the sum in quadrature of these three uncertainties. The scale uncertainty dominates the total uncertainty in the cross sections. The total uncertainty decreases with \etg for the \yb and \yc cross sections, from around 25\% to around 15\% in the measured range. These uncertainties are also evaluated for the ratio of the cross section in the central photon pseudorapidity region to that in the forward region by separately propagating each uncertainty variation to the ratio, assuming full correlations between the two regions. The uncertainties are then assessed in a similar way as those in the cross sections. The total uncertainty in the cross-section ratios is nearly constant with \etg and is about 5\%. The scale uncertainty dominates the total uncertainty in the cross-section ratios, except in the case of the predictions using the fitted charm PDF set from NNPDF3.1nlo, for which the PDF uncertainty dominates. The total uncertainty in the cross-section ratios using this fitted charm PDF increases with \etg, from about 5\% at 25~\GeV\ to about 15\% at 350~\GeV.  No uncertainties are assessed for the LO predictions as the scale uncertainties are expected to be large and unreliable.

\section{Event selection and calibration}
\label{sec:Selection}

This measurement makes use of the full dataset of \pp collisions at a centre-of-mass energy of 8~\TeV, recorded by the ATLAS detector in 2012. Only events taken during stable beam conditions when the ATLAS detector operation satisfied data-quality conditions are considered. Single-photon triggers with \etg thresholds of 20, 40, 60, 80, 100 and 120~\GeV, which have efficiencies measured to be greater than 99\% with respect to the offline selection requirements, were used to record events. Below 125~\GeV, each bin of the measurement is populated by a single trigger, while the remaining bins are all populated by the highest threshold trigger. Due to their higher rates and considering the available bandwidth, all but the 120~\GeV\ trigger were prescaled, such that only some of the events satisfying the trigger requirement were recorded. Events recorded by a prescaled trigger are weighted by the ratio of the unprescaled recorded luminosity to the recorded luminosity of the respective trigger. The integrated luminosity of the data ranges from $4.58 \pm 0.09$ pb$^{-1}$ for the 20~\GeV\ trigger to $20.2 \pm 0.4$~fb$^{-1}$ for the unprescaled 120~\GeV\ trigger~\cite{Lumi}.

Events are required to have a hard reconstructed primary vertex consistent with the nominal interaction point and at least two associated tracks with transverse momentum, $\pt$, greater than $400$~\MeV. In events where more than a single vertex satisfies these criteria, the vertex with the highest $\sum \pt^2$ of associated tracks is considered as the hard vertex. The dataset exhibits an average of 19 \pp interactions per bunch crossing, where the interactions not associated to the hard vertex are referred to as \emph{pile-up}, an effect that is taken into account in the reconstruction. Effects related to events containing more than one hard vertex are estimated to be negligible, below the percent level, and are not considered.

Detector-level photon candidates are built from ECAL cell clusters with transverse energies greater than 2.5~\GeV. They fall into two categories: unconverted and converted photons. Unconverted photons have no tracks associated with the cluster. Converted photons have associated tracks that are consistent with the signature of a photon interacting upstream. The overall photon reconstruction efficiency is 96\% for prompt photons with $\etg > 25$~\GeV\ \cite{egammaEff}. Converted and unconverted photons are calibrated separately, making use of both the calorimeter and the tracking information to correct the calorimeter response for upstream energy losses and leakage \cite{egammaCal}. In the simulation, only detector-level photons that match a particle-level prompt photon using a cone of $\Delta R = 0.2$ are considered. 
Detector-level photon candidates are subject to a two-stage shower-shape-based identification criterion. The first stage scrutinizes the leakage into the hadronic calorimeter, the lateral size and shape of the cluster in the second ECAL layer and the shower width in the first ECAL strip layer \cite{egammaEff}. The second stage imposes additional criteria that are sensitive to the lateral shape of the shower in the ECAL strip layers, providing discrimination against neutral hadron decays into pairs of photons. 
This stage can be inverted to populate background enriched sideband regions.
These two stages, which together are referred to as the \emph{tight} criteria, are used in previous ATLAS prompt-photon analyses \cite{STDM-2011-28,diphoton7TeV2,yJet7TeV,incphoton7TeV,incphoton8TeV,yJet8TeV,incphoton13TeV,diphoton8TeV}. 
In the simulation, the distributions of the shower-shape variables used for photon identification are corrected to reproduce those observed in the data. Further, event weights are applied to simulated events whose leading photon satisfies the tight selection criteria such that the identification efficiency matches that of the data for both the converted and unconverted photons. These event weights are typically within 3\% of unity.

Both the detector-level and particle-level photon candidates are required to exhibit an isolated signal, a requirement that targets prompt-photon production and discriminates against jets misidentified as photons. 
This criterion is imposed through the definition of the calorimeter isolation variable \etiso. This variable is defined at the detector level as the sum of the transverse energies recorded in clusters within a distance of $\Delta R = 0.4$ around the photon, excluding the contributions in a fixed-size window centred on the photon candidate of size $0.125 \times 0.1715$ in $\Delta\eta \times \Delta\phi$. The variable is then corrected for contributions from the pile-up and the underlying-event \cite{incphoton7TeV}. In the simulations, corrections are also applied to account for mismodelling of the mean and the spread of the detector-level \etiso distribution. These corrections are derived by matching the simulated \etiso distribution to the signal photon \etiso distribution extracted from data using a data-driven template fit. 
In the simulation, the particle-level \etiso is calculated by adding the transverse energy of all particles with a lifetime greater than 10~ps within a distance of $\Delta R = 0.4$ around the photon. Muons and neutrinos, however, are excluded since they deposit little or no energy in the calorimeter. The \etiso is then corrected for the energy density of the underlying event \cite{JetAreas}.
A sliding \etiso requirement is used to impose the isolation criterion at both the detector level and the particle level: $\etiso < 4.8~\GeV + 0.0042 \times \etg$. The \etg dependent nature of the requirement improves the acceptance of high-\etg signal photons, yielding a roughly constant 92\% signal efficiency. 
The efficiency for isolated detector-level unconverted (converted) photons in the data to satisfy the tight criteria is approximately 75\% (75\%) for an \etg of 25 \GeV\ and 95\% (98\%) for 400~\GeV\ \cite{egammaEff}. The \etiso requirement is inverted with a 2~\GeV\ gap, $\etiso > (4.8+2)~\GeV + 0.0042 \times \etg$, to populate regions used for the photon background sideband subtraction. The 2~\GeV\ gap is used to reduce the amount of signal in the sideband regions.

Detector-level jets are built using the \antikt algorithm \cite{antiKt}, implemented in the FastJet package \cite{fastjet}, taking as input calibrated topological clusters \cite{TopoClusters} in the calorimeter and a parameter of $R=0.4$. The jets undergo a five-stage sequential calibration \cite{JGSC}. This calibration includes corrections based on the cluster shape and location, the jet area and pile-up \cite{JetAreas}, the response of simulated particle-level jets, the combined shower-structure and tracking information and finally the data-driven $\gamma$+jet, $Z$+jet and multijet $p_\mathrm{T}$-balance of the energy scale \cite{JES}. Detector-level jets are required to satisfy quality criteria that ensure they are not affected by, or are the result of, detector defects and noise, cosmic rays or non-collision beam-related backgrounds \cite{JES,JES7TeV}. To reduce the impact of jets coming from pile-up interactions, detector-level jets with $\ptj < 50$~\GeV\ and $|\etaj| < 2.4$ are also required to have at least 50\% of the momentum of associated tracks to originate from the hard vertex \cite{Aad:2015ina}. Simulated particle-level jets used in the analysis are built using the \antikt algorithm taking as input all particles with a lifetime greater than 10~ps and a radius parameter of $R=0.4$.

Simulated particle-level and detector-level jets are assigned a flavour based on the following hadron matching scheme. If a $b$-hadron with $\pt > 5$~\GeV\ is found within $\Delta R= 0.3$ of a jet then it is considered to be a $b$-jet. If a jet that is not a $b$-jet is found to have a $c$-hadron with $\pt > 5$ ~\GeV\ within $\Delta R= 0.3$ then it is considered to be a $c$-jet. If a jet that is not a $b$- or a $c$-jet is found to have a $\tau$-lepton with $\pt > 5$~\GeV\ within $\Delta R= 0.3$ then it is considered to be a $\tau$-jet. If a jet is found to be neither a $b$-jet, a $c$-jet nor a $\tau$-jet it is considered to be a light jet. The contribution of $\tau$-jets in the measurement is negligible.

At the detector level, only the highest-\et (leading) photon that satisfies the first stage of the photon identification criteria is considered. The photon candidate is then required to have $\etg > 25$~\GeV\ and $|\eta^{\gamma}| < 2.37$, excluding the transition region between the barrel and endcap ECAL modules $1.37 < |\etag| < 1.56$. 
The photon candidate must then satisfy the second stage of the identification criteria. If it fails, it is instead used to populate regions used for the photon background sideband subtraction. 
The photon candidate must then satisfy the \etiso criterion. If it fails, but satisfies the inverted \etiso criterion, it is instead used to populate regions used for the photon background sideband subtraction. 
Next, only the leading jet with $\Delta R > 0.4$ from the photon candidate is considered. This jet is required to have $\pt^{\text{jet}} > 20$~\GeV, $|\eta^{\text{jet}}|<2.5$ and to be separated from the photon candidate by $\Delta R > 1$. This last angular separation requirement ensures that the measured signals of the leading jet and the leading photon do not overlap. 

At the particle level, only the leading photon is considered. The fiducial requirements imposed at particle level are similar to those used at the detector level, but using the jet rapidity instead of its pseudorapidity, and are summarized in Table \ref{tab:fiducial}. 

\begin{table}[t]
\centering
\small
\caption{Particle-level requirements defining the fiducial region. The determination of the jet flavour and the calculation of \etiso are described in the text.}
\label{tab:fiducial}
\begin{tabular}{ l  c  c }
\toprule
Particle-level selection &Leading $\gamma$&Leading jet with $\Delta R^{\gamma-\text{jet}} > 0.4$\\ 
\midrule
Transverse momentum & $\etg > 25~\GeV$ & $\ptj > 20~\GeV$\\
Rapidity & $|\etag| < 1.37$ or $1.56 < |\etag| < 2.37$ & $|y^{\text{jet}}| < 2.5$\\
Isolation & $\etiso < 4.8~\GeV + 0.0042 \times \etg$ & --- \\
\cmidrule{2-3}
Angular separation & \multicolumn{2}{c}{ $\sqrt{(\Delta y) ^2 + (\Delta \phi)^2}> 1$ }\\
\bottomrule
\end{tabular}
\end{table}

Detector-level jets are assigned a $b$-tagging discriminant value by the MV1c algorithm.
The MV1c algorithm is a neural network that takes as input the discriminants of three tagging algorithms, analogous to the MV1 algorithm \cite{bJetCal}, but is trained to identify $b$-jets with enhanced rejection of $c$-jets. The three tagging algorithms input to the MV1c tagger are based on different aspects of jet tracking information that are sensitive to the presence of secondary vertices originating from HF decays: the IP3D algorithm is sensitive to the displacement of the tracks associated to the jet from the primary vertex, the SV1 algorithm reconstructs secondary vertices and the JetFitter algorithm is sensitive to secondary and tertiary vertices that are kinematically consistent with the decay chain of a $b$- or $c$-hadron. The MV1c tagger discriminant distribution is divided into five bins delimited by four cuts corresponding to the $b$-jet identification efficiencies in simulated top quark pair ($t\bar{t}$) events of 80\%, 70\%, 60\% and 50\%, and bounded by the trivial 100\% and 0\% cut values. The discriminant distribution in the simulation is calibrated using event weights according to the jet flavour and kinematics, such that the overall efficiency of each cut value in the simulation matches that of the data. This calibration considers the correlations between the discriminant bins and has been used in a prior ATLAS measurement~\cite{ttbar}. 
The efficiency of these cuts in simulated events satisfying the $\gamma$+jet selection used for this measurement is typically 2--5\% lower than that measured in the $t\bar{t}$ calibration analysis. For the event weights, since they are ratios, this difference is mostly cancelled. No statistically significant difference is expected between the event weights in $\gamma$+jet and in $t\bar{t}$. These event weights deviate from unity by up to 30\%.

Separate $b$-, $c$- and light-jet calibrations are used to correct the efficiency of the discriminant cuts to better match the data. The $b$-jet calibration~\cite{bjetcalib} uses an unbinned maximum-likelihood fit of simulated templates to extract the $b$-jet tagging efficiency distribution in data using a $t\bar{t}$ selection, which has a high $b$-jet purity. The fit considers the individual probability for each jet in a given event to be tagged, thereby exploiting per-event jet-flavour correlations. The $c$-jet calibration~\cite{cljetcalib} uses a sample of reconstructed $D^{*\pm}$ mesons to extract the $c$-jet tagging efficiency by fitting simultaneously the $D^{*\pm}$ yield with and without applying a cut on the MV1c discriminant. As $b$-hadrons can also produce $D^{*\pm}$, a fit of the $D^{0}$ pseudo-proper lifetime is used to subtract the $b$-jet background. The inclusive $c$-jet tagging efficiency is then derived using existing dedicated decay measurements~\cite{Patrignani:2016xqp} and simulations based on \textsc{EvtGen}~\cite{EvtGen} to extrapolate it from the measured $D^{*\pm}$ $c$-jet efficiency. Light jets can be tagged as $b$-jets mainly due to the finite tracking resolution. The light-jet calibration~\cite{cljetcalib} involves inverting the sign of some of the criteria imposed on the impact parameter and the decay-length significance in the MV1c algorithm. The resulting discriminant distribution for all flavours is similar to that of the nominal MV1c discriminant distribution for light jets. Consequently, the jet tagging efficiency obtained using this method is taken as the light-jet tagging efficiency after correcting it for the effects of HF jets, long-lived particles and material interactions.

\section{Signal extraction}
\label{sec:Background}

A data-driven two-dimensional sideband technique \cite{STDM-2011-28,yJet7TeV,incphoton7TeV,incphoton8TeV,yJet8TeV,incphoton13TeV} is applied to estimate and subtract photon background contamination from the data yield in each MV1c tagger discriminant bin, in every bin of the measurement. By this means, any correlation between jets misidentified as photons and the flavour of the accompanying jet is taken into account. The technique relies on the use of three background-dominated control regions: two of them created by individually inverting separate aspects of the signal photon selection criteria and a third region created by inverting these two aspects simultaneously. The first aspect is the inversion of the second stage of the photon identification criteria based on shower shapes. The second aspect is the inversion of the $\etiso$ selection requirement. Both of these aspects provide discrimination against jets faking photons and photons arising from hadron decays. The three background regions in the data are then used to estimate the prompt photon yield in the signal region, taking into account the estimated leakage of signal into these regions using the simulations. The procedure hinges on the assumption that the two inverted aspects of the selection criteria are uncorrelated for background events. Deviations from this assumption are small and are taken into account as an uncertainty. The photon purity, i.e.\ the fraction of signal photons, is typically $55$\% in the lowest bin of $\etg$, rising steadily to greater than $95$\% around $400$~\GeV. 
The largest correlation between the photon purity and the MV1c discriminant is observed in the 25--45~\GeV\ bin of $\etg$ in the central region where the photon purity exhibits a relative increase of roughly 15\% from the 100--80\% MV1c $b$-jet efficiency bin to the 50--0\% bin.

\begin{figure}[htbp]
    \centering
    \begin{subfigure}{0.49\textwidth}
        \includegraphics[width=\textwidth]{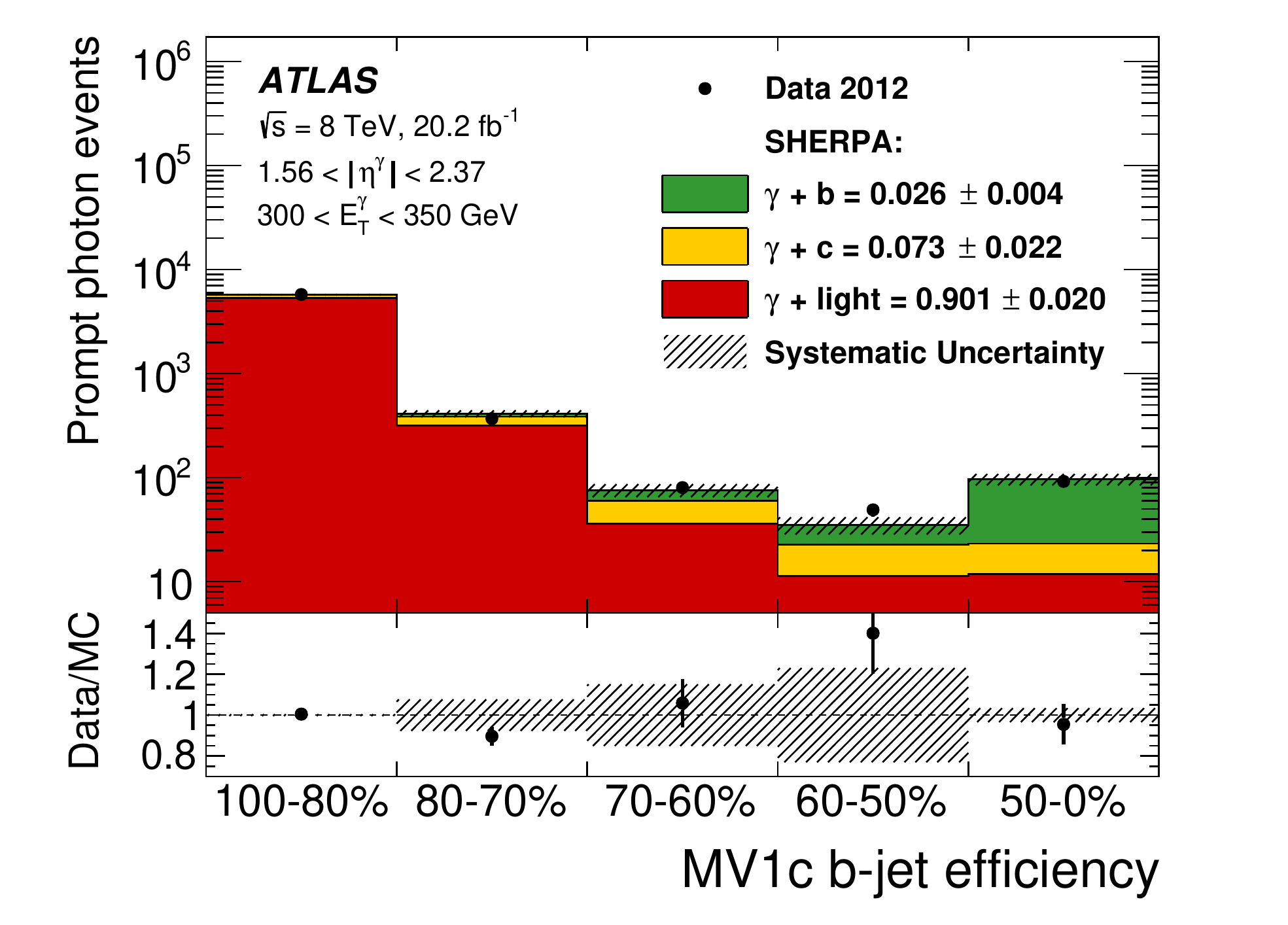}
        \caption{}
        \label{fig:fita}
    \end{subfigure}
    \begin{subfigure}{0.49\textwidth}
        \includegraphics[width=\textwidth]{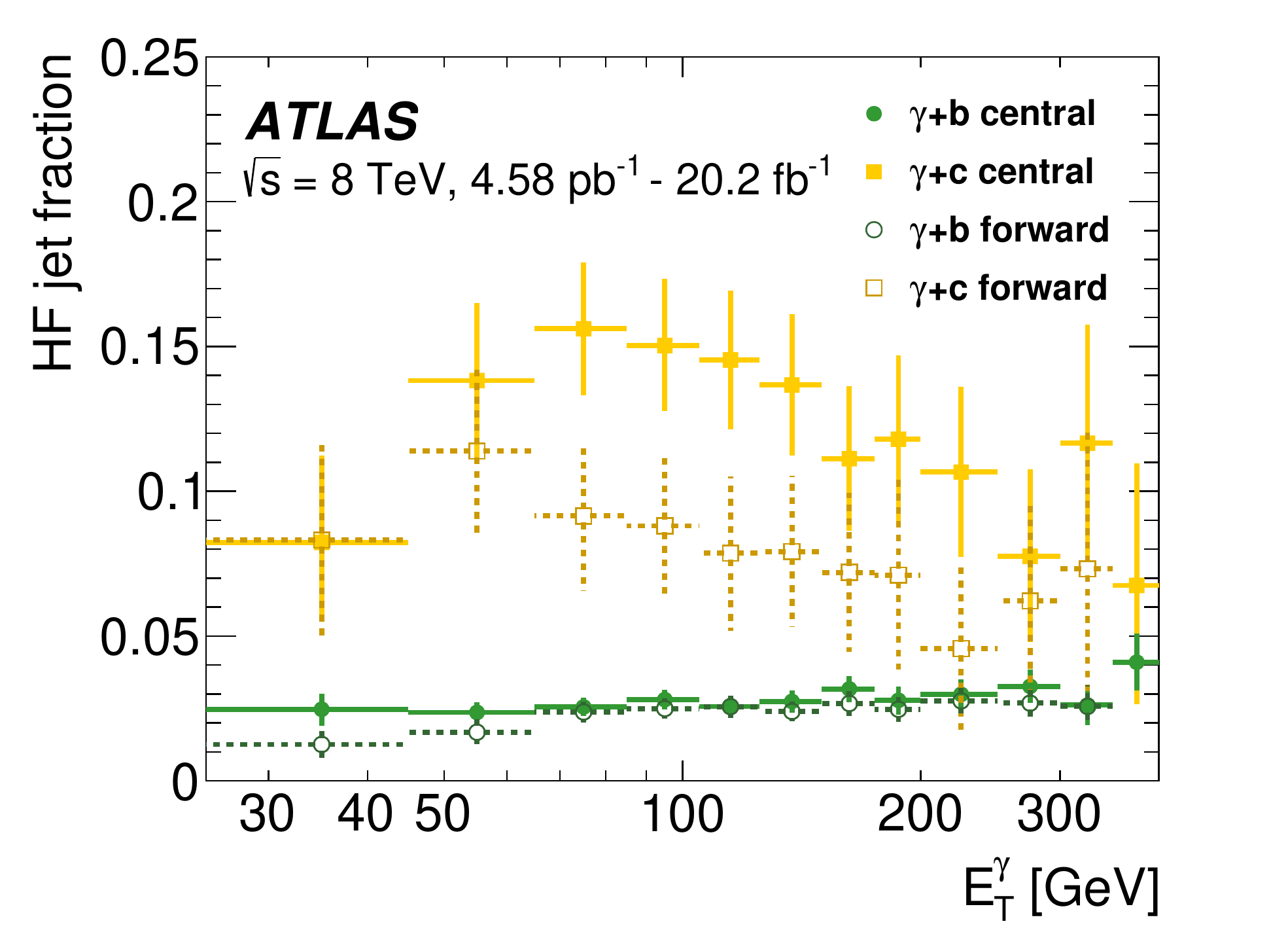}
        \caption{}
        \label{fig:fitb}
    \end{subfigure}
    \caption{(a) Example of a template fit to the MV1c tagger discriminant distribution used to measure the $\gamma+b$ and $\gamma+c$ fractions. The data yield is shown after subtraction of background photons. The error bars on the data correspond to the statistical uncertainty. The systematic uncertainty band takes into account the uncertainty correlations between the data and the MC templates. The numbers in the legend are the fractions of each template after the fit and their statistical uncertainties. (b) The heavy-flavour jet fractions obtained from the template fits as a function of the photon transverse energy, \etg. The fractions are relative to the yield of selected $\gamma+\text{jet}$ data events after subtraction of background photons. The error bars correspond to the total uncertainty, including the statistical uncertainty and the complete set of systematic uncertainties. The central and forward regions are defined respectively as $|\etag|<1.37$ and $1.56<|\etag|<2.37$.}
    \label{fig:fit}
\end{figure}

Following the photon background subtraction, the MV1c tagger distribution of the signal photon yield is used to extract the $b$-jet and $c$-jet fractions in each bin of the measurement. Simulated discriminant shapes for $b$-jets, $c$-jets and light jets, which are corrected using factors derived from the aforementioned tagging calibration analyses, are used to perform a template fit. The shape uncertainty in the simulated templates is derived from the uncertainties in the tagging calibration, taking into account correlations between the discriminant bins.
The template fit is performed as a binned maximum-likelihood fit. Figure \ref{fig:fita} shows the template fit for the 300--350~\GeV\ bin of \etg in the forward region, which is particularly sensitive to intrinsic charm. The quality of this fit is similar to that of the others. The general features are that light jets populate the high $b$-jet efficiency side of the discriminant, $b$-jets populate the low $b$-jet efficiency side and $c$-jets lie between the two. Since the jet-flavour fractions are measured simultaneously, they are correlated in each bin of the measurement. Figure \ref{fig:fitb} shows the measured $b$- and $c$-jet fractions for central and forward photons.
The fraction of $c$-jets displays a maximum between $50$ and $80~\GeV$ while that of $b$-jets displays a slight monotonic increase. These features are predicted in the particle-level simulations.

\section{Cross-section measurement procedure}
\label{sec:Procedure}

The following equation outlines the procedure, making use of the \SHERPA simulation, used to compute the cross section from the data yield:
\begin{linenomath}
  \begin{equation*}
    \left(\frac{\mathrm{d}\sigma^{\gamma+\text{HF-jet}}}{\mathrm{d}E_\mathrm{T}^{\gamma}}\right)_i = \frac{1}{(\Delta E_\mathrm{T}^\gamma)_i}\frac{1}{(\mathcal{L_\text{int}^\text{trig}})_i}\frac{1}{\epsilon^\text{trig}_i} C_i f_i^\text{HF-jet}\sum_{j\in\text{MV1c}}p_{ij}^\gamma N_{ij}^{\text{Data}}.
  \end{equation*}
\end{linenomath}
The left side of this equation is the measured cross section 
corrected back to the particle level in bin $i$ of $\etg$, $(\Delta E_T^\gamma)_i$ is the bin width, $(\mathcal{L_\text{int}^\text{trig}})_i$ is the integrated luminosity of the trigger, $\epsilon^\text{trig}_i$ is the trigger 
efficiency, $C_i$ is the particle-level correction factor, $f_i^\text{HF-jet}$ 
is the measured HF-jet fraction, $p_{ij}^\gamma$ is the measured signal-photon purity in a tagger 
discriminant bin $j$ and $N_{ij}^{\text{Data}}$ is the yield of selected $\gamma+\text{jet}$ data events. The particle-level 
correction factor accounts for detector effects, including the detector resolution and the signal 
reconstruction efficiency, using the one-dimensional bin-by-bin approach, yielding a measurement that is directly 
comparable to other experimental results and theoretical predictions. 
The bin-by-bin approach, used in previous ATLAS photon results \cite{STDM-2011-28,yJet7TeV,incphoton7TeV,incphoton8TeV,yJet8TeV,incphoton13TeV}, 
uses factors defined as the ratio of the particle-level to the detector-level \etg distributions 
derived using the simulation for each bin of the measurement:
\begin{linenomath}
\begin{equation*}
C_i= \frac{N^{\text{particle,~}\gamma+\text{HF-jet}}_{i}}{N^{\text{detector,~}\gamma+\text{HF-jet}}_{i}}.
\end{equation*}
\end{linenomath}
The accuracy of this approach relies on the detector-level bin-migration effects being well described by the simulation since correlations between adjacent bins are neglected. As the cross sections are measured differentially with respect to \etg, this condition is met since the \etg resolution is much smaller than the bin width. In the central region, migrations are less than 5\%, while in the forward region they are less than 10\%. 
The values of the correction factors decrease with \etg, driven by the improving photon identification efficiency, from typically $1.9$ ($1.7$) at $25$~\GeV\ to $1.2$ ($1.2$) at $400$~\GeV\ for \yb (\yc) events. They do not have a strong dependence on $|\etag|$.

\section{Measurement uncertainties}
\label{sec:Uncertainties}

Uncertainties affecting the measurement which originate from the finite numbers of data and MC events are considered together with systematic uncertainties related to the detector calibration and analysis techniques. The bootstrap resampling technique~\cite{BootStrap} is used to assess the statistical uncertainties by creating an ensemble of statistically equivalent measurements using event weights, randomly chosen for each event from a Poisson distribution with a mean of one, applied to either the data or MC events. The 68\% confidence interval of the distribution of these measurements is taken as the statistical uncertainty. The systematic uncertainties are derived by varying a parameter in the simulated events, repeating the complete analysis with this varied parameter and taking the difference between the new measured value and the nominal measurement as the uncertainty. The bootstrap resampling technique is then used to evaluate the statistical uncertainty in each systematic variation. Variations that are not statistically significant undergo a bin-merging procedure over an increasing number of \etg bins to improve their significance. Following this procedure, only statistically significant variations are considered as systematic uncertainties. This procedure, however, gives rise to an interplay between the MC statistical uncertainty and the systematic uncertainties.

Sources related to the detector calibration include the photon energy scale and resolution~\cite{egammaCal}, the photon identification efficiency~\cite{egammaEff}, the jet energy scale and resolution~\cite{JES}, the inefficiency of the pile-up jet removal cut~\cite{Aad:2015ina} and the MV1c tagger discriminant for the three jet flavours~\cite{bjetcalib,cljetcalib}. The energy scale and resolution of the photon and the jet have several uncertainty components that encompass both the imperfect knowledge of the detector response and the analysis techniques used to derive the calibration. The calibration is varied according to its uncertainties to assess the impact on the measurement. The calibrations of the photon identification efficiency and of the MV1c tagger discriminant have uncertainties related to the analysis techniques in which they were derived. For the case of the MV1c discriminant, the uncertainties are mostly related to the modelling of the track multiplicity and the misidentification of the hadron flavours. These factors are varied according to their uncertainties. However, some of their uncertainty components, such as those of the jet energy scale and resolution, are correlated with those of this measurement. As such, these components are varied coherently both in the discriminant calibration and in this analysis. To assess the uncertainty due to the pile-up jet removal cut, the 50\% requirement on track momentum from the hard vertex is varied to 53\% and 47\%. The magnitude of the variation is motivated by the cut efficiency difference between the data and the simulation.

The uncertainties related to the analysis techniques are similar to those in the ATLAS inclusive photon and $\gamma+$jet analyses at 8~\TeV~\cite{incphoton8TeV,yJet8TeV}. Specifically, the assumption that the photon background regions are uncorrelated in the two-dimensional sideband method is assessed by varying the correlation by 10\%. The magnitude of this variation corresponds to the size of the measured correlation in control regions of the data. The two definitions of the background regions in the sideband method are varied as follows. The photon identification reversal is varied by adding, or by removing, an identification criterion based on the first layer of the calorimeter. The inverted photon isolation energy cut is increased and decreased by 2~\GeV, motivated by the difference seen in the isolation energy resolution between data and the simulations. An uncertainty related to the photon isolation energy corrections is obtained by varying them according to the differences seen between \SHERPA and \PYTHIA. A prompt-photon modelling uncertainty is assessed by varying the relative fraction of hard-scatter photons and radiated photons generated in \PYTHIA. Similarly, the change in the measurement when using simulated samples from \PYTHIA instead of \SHERPA is taken as an uncertainty, assessing the differences between the non-perturbative QCD models used by the generators. Possible migration effects in the bin-by-bin particle-level correction factors are also taken as an uncertainty.

The uncertainties in the cross-section ratios are obtained by propagating the individual systematic variations of the central and forward cross sections to the ratio and taking the resulting variations as the uncertainties. As most systematic uncertainties are positively correlated between these two pseudorapidity regions, their effect on the ratios is smaller than on the cross sections. However, some components of the photon energy scale and of the light-jet tagging efficiency uncertainties are uncorrelated between the two regions. As the two regions considered are exclusive in $|\etag|$, the data and MC statistical uncertainties in the ratio exceeds those in the cross sections.
  
The ranges of the size of the relative uncertainties in the measurement as a function of \etg due to the various sources are provided in Table~\ref{tab:unc}. The dominant uncertainties are due to the finite number of data events and the calibration of the MV1c tagger discriminant. This latter source of uncertainty primarily impacts the measurement through the HF jet fraction determined in the MV1c template fit. Uncertainties arising from the photon energy resolution, the data-driven photon isolation energy corrections and the pile-up jet removal cut are negligible and not listed in Table~\ref{tab:unc}, nor are they considered further. The remaining uncertainties are added in quadrature to give the total uncertainty. The total uncertainty is largest at low and high \etg, reaching a minimum at about 100~\GeV. At higher \etg values than those measured, the total relative uncertainty becomes excessively large and the MV1c template fit becomes unstable, curtailing the reach of the measurement. The total relative uncertainty is larger in the forward region than in the central region due to the higher data and MC statistical uncertainties. The \yc measurement is affected by larger relative uncertainties than the \yb measurement since the MV1c tagger discriminates better $b$-jets than $c$-jets.

\begin{table}[t]
\centering
\footnotesize
\caption{Range of the size of the relative uncertainties in the measured cross sections along \etg for the different uncertainty sources. The central region refers to the cross sections in the range $|\etag|<1.37$, the forward region refers to the cross sections in the range $1.56<|\etag|<2.37$ and the ratio refers to the ratio of the cross section in the central region to that in the forward region. The systematic variations must be statistically significant to be considered as systematic uncertainties. Uncertainties with values listed as $<0.1$ are not statistically significant.}
\label{tab:unc}
\begin{tabular}{l c c c c c c}
\toprule
&\multicolumn{6}{c}{Uncertainty [\%]}\\
\cmidrule{2-7}
Uncertainty source&\multicolumn{3}{c}{$\gamma+b$}&\multicolumn{3}{c}{$\gamma+c$}\\
\cmidrule(r){2-4}
\cmidrule(l){5-7}
&Central&Forward&Ratio&Central&Forward&Ratio\\
\midrule
MC statistical uncertainty&$1.9-\phantom{0}6.4$ & $3.1-14\phantom{.0}$ & $3.6-17\phantom{.0}$ & $\phantom{0}2.5-24\phantom{.0}$ & $\phantom{0}6.0-33\phantom{.0}$ & $6.1-39\phantom{.0}$\\
Photon energy scale&$0.2-\phantom{0}2.5$&$0.7-\phantom{0}5.3$&$0.9-\phantom{0}1.9$&$\phantom{0}0.2-\phantom{0}1.0$&$\phantom{0}0.0-\phantom{0}0.2$&$0.5$\\
Photon identification efficiency&$0.2-\phantom{0}1.2$&$0.4-\phantom{0}1.8$&$0.1-\phantom{0}0.5$&$\phantom{0}0.2-\phantom{0}1.3$&$\phantom{0}0.4-\phantom{0}1.7$&$0.1-\phantom{0}0.5$\\
Jet energy scale&$0.6-\phantom{0}4.8$&$0.7-\phantom{0}4.6$&$0.1-\phantom{0}0.2$&$\phantom{0}0.2-\phantom{0}2.3$&$\phantom{0}0.2-\phantom{0}2.8$&$0.1-\phantom{0}0.5$\\
Jet energy resolution&$0.0-\phantom{0}2.4$&$0.0-\phantom{0}1.0$&$0.0-\phantom{0}0.1$&$\phantom{0}0.0-16\phantom{.0}$&$\phantom{0}0.2-\phantom{0}5.7$&$0.4-\phantom{0}2.5$\\
\bj tagging efficiency&$2.4-17\phantom{.0}$&$2.5-15\phantom{.0}$&$0.1-\phantom{0}0.6$&$\phantom{0}0.4-12\phantom{.0}$&$\phantom{0}0.5-\phantom{0}8.3$&$0.2-\phantom{0}2.3$\\
\cj tagging efficiency&$5.7-18\phantom{.0}$&$5.3-11\phantom{.0}$&$2.3-\phantom{0}6.9$&$\phantom{0}6.0-18\phantom{.0}$&$\phantom{0}6.4-18\phantom{.0}$&$0.4-\phantom{0}2.7$\\
Light-jet tagging efficiency&$4.9-15\phantom{.0}$&$6.1-31\phantom{.0}$&$1.6-\phantom{0}8.3$&$12\phantom{.0}-46\phantom{.0}$&$21\phantom{.0}-57\phantom{.0}$&$8.4-28\phantom{.0}$\\
Sideband definition&$0.2-\phantom{0}3.0$&$0.2-\phantom{0}2.9$&$0.1-\phantom{0}0.8$&$\phantom{0}0.2-\phantom{0}3.4$&$\phantom{0}0.2-\phantom{0}1.2$&$0.1-\phantom{0}0.6$\\
Sideband correlation&$0.2-\phantom{0}4.5$&$0.4-13\phantom{.0}$&$0.2-10\phantom{.0}$&$\phantom{0}0.2-\phantom{0}4.2$&$\phantom{0}0.5-\phantom{0}5.2$&$0.3-\phantom{0}1.2$\\
Prompt-photon modelling&$2.2-\phantom{0}2.5$&$2.4$&$4.2-\phantom{0}6.7$&$\phantom{0}1.5-\phantom{0}2.8$&$<0.1$&$<0.1$\\
Non-perturbative QCD models&$2.3$&$7.3$&$11$&$<0.1$&$<0.1$&$<0.1$\\
Particle-level migration effects&$0.8-\phantom{0}2.9$&$0.4$&$1.2-\phantom{0}4.3$&$\phantom{0}0.9-\phantom{0}3.1$&$<0.1$&$0.6-\phantom{0}3.0$\\
Luminosity&$1.9$&$1.9$&$0$&$1.9$&$1.9$&$0$\\
\midrule
Total systematic uncertainty&$12\phantom{.0}-25$&$13\phantom{.0}-38$&$14\phantom{.0}-22$&$15\phantom{.0}-56$&$25\phantom{.0}-61$&$11\phantom{.0}-48$\\
Data statistical uncertainty&$\phantom{0}1.5-13$&$\phantom{0}2.1-37$&$\phantom{0}2.5-58$&$\phantom{0}1.1-27$&$\phantom{0}2.9-33$&$\phantom{0}3.2-47$\\ 
\midrule
Total uncertainty&$13-27$&$14-54$&$14-62$&$15-62$&$26-66$&$14-66$\\
\bottomrule
\end{tabular}
\end{table}

\section{Results}
\label{sec:Results}
 
\begin{figure}[htbp]
    \centering
    \begin{subfigure}{0.49\textwidth}
        \includegraphics[width=\textwidth]{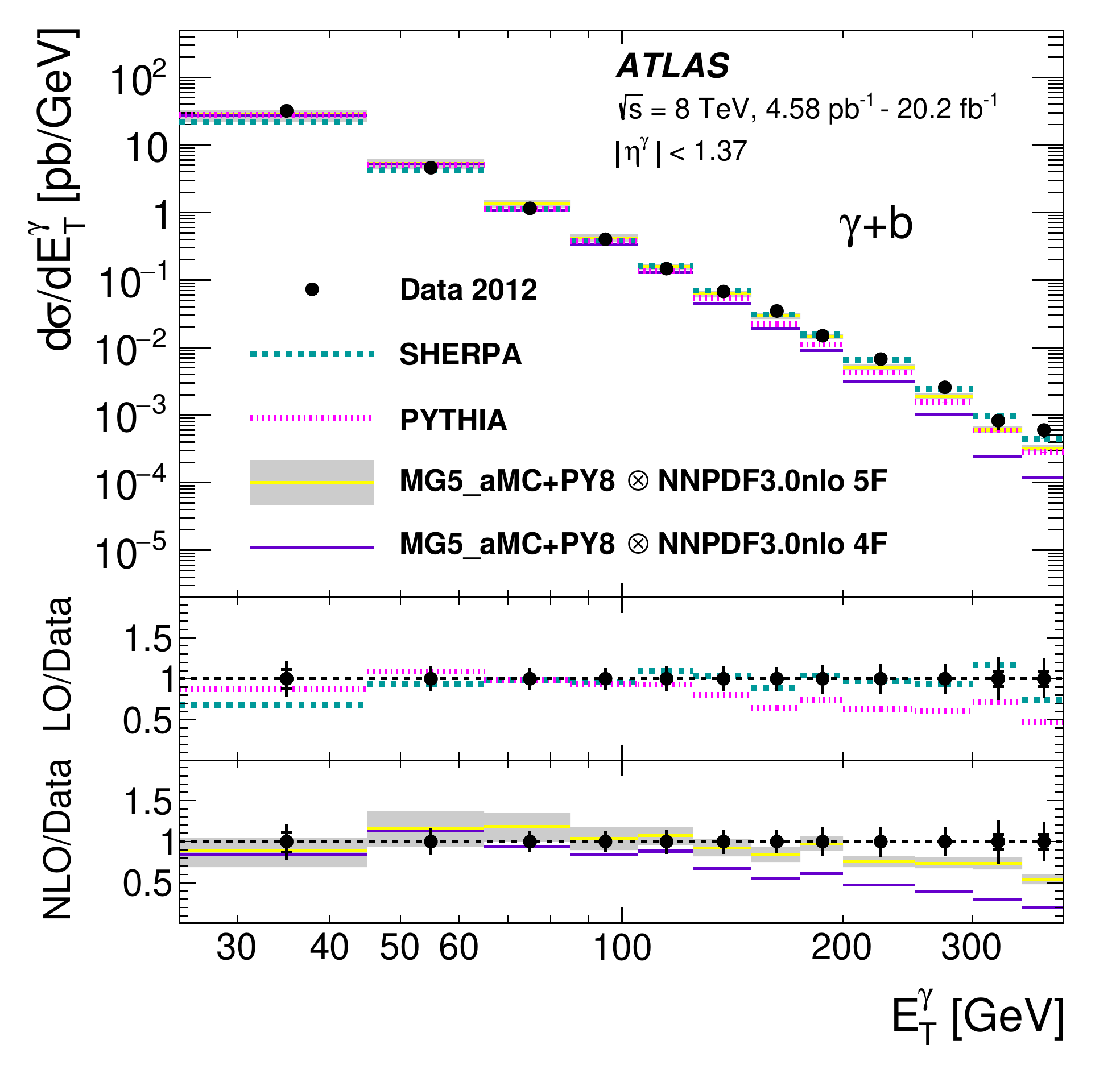}
        \caption{}
        \label{fig:xseca}
    \end{subfigure}
    \begin{subfigure}{0.49\textwidth}
        \includegraphics[width=\textwidth]{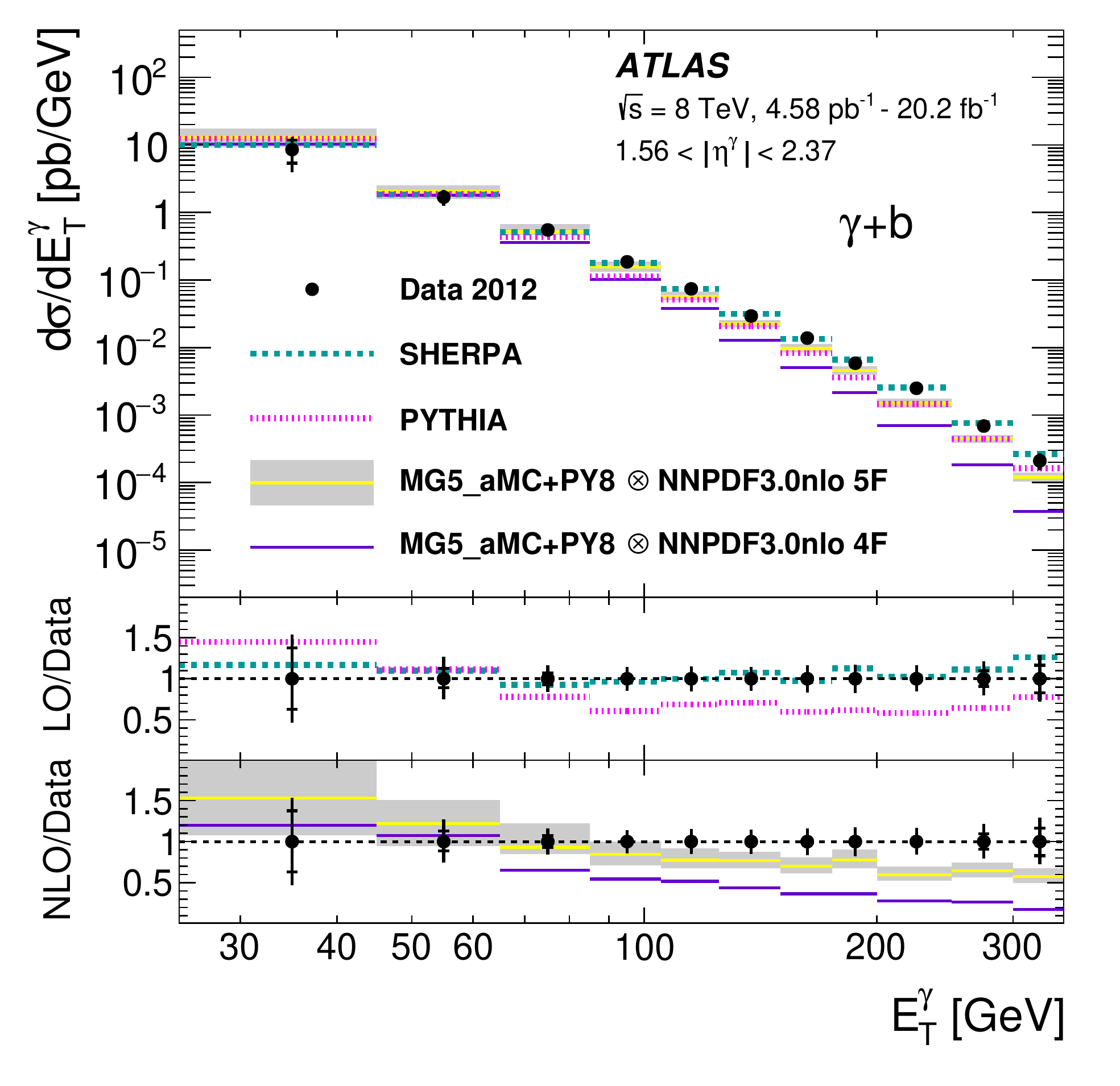}
        \caption{}
        \label{fig:xsecb}
    \end{subfigure}
        \begin{subfigure}{0.49\textwidth}
        \includegraphics[width=\textwidth]{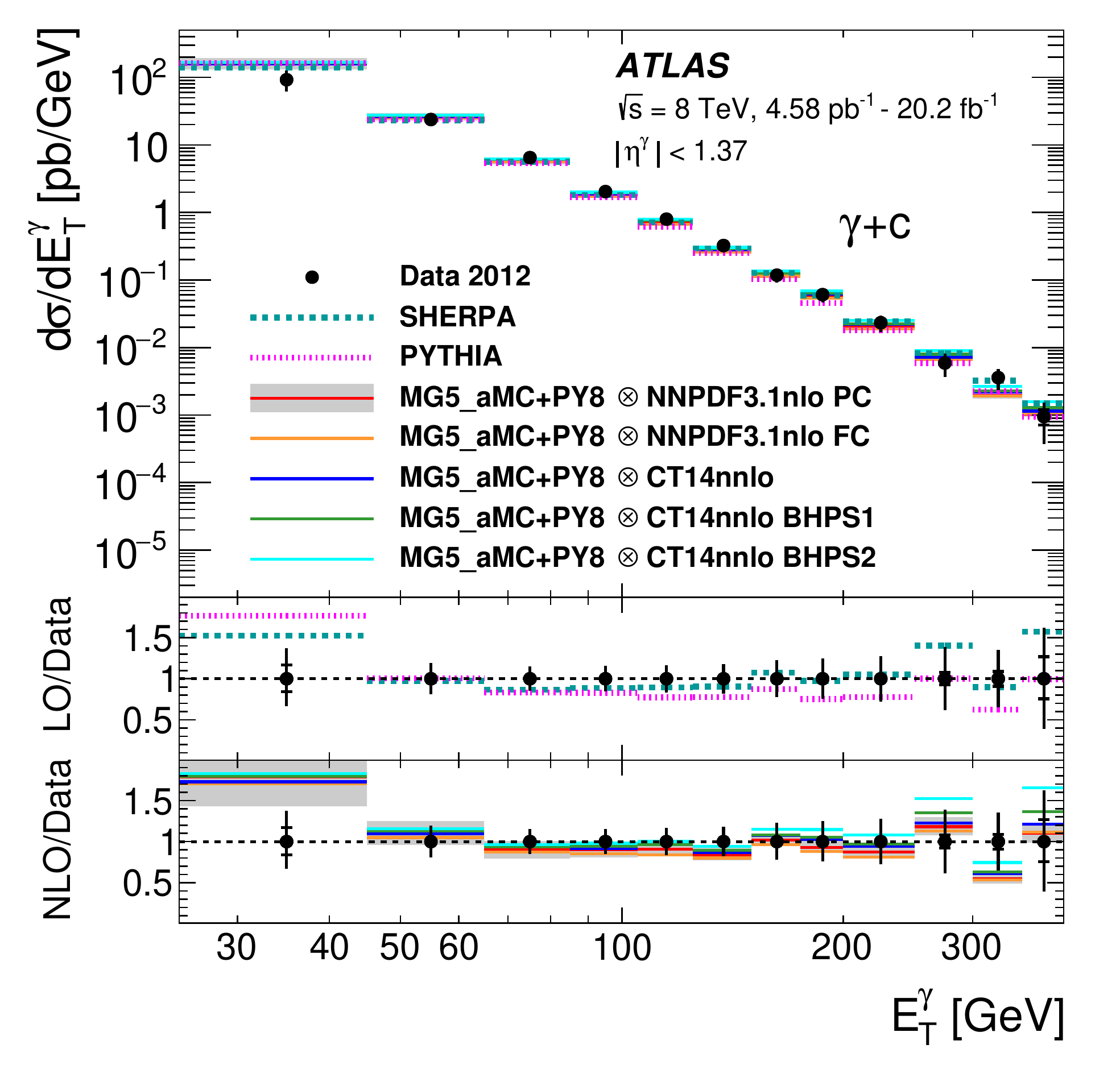}
        \caption{}
        \label{fig:xsecc}
    \end{subfigure}
    \begin{subfigure}{0.49\textwidth}
        \includegraphics[width=\textwidth]{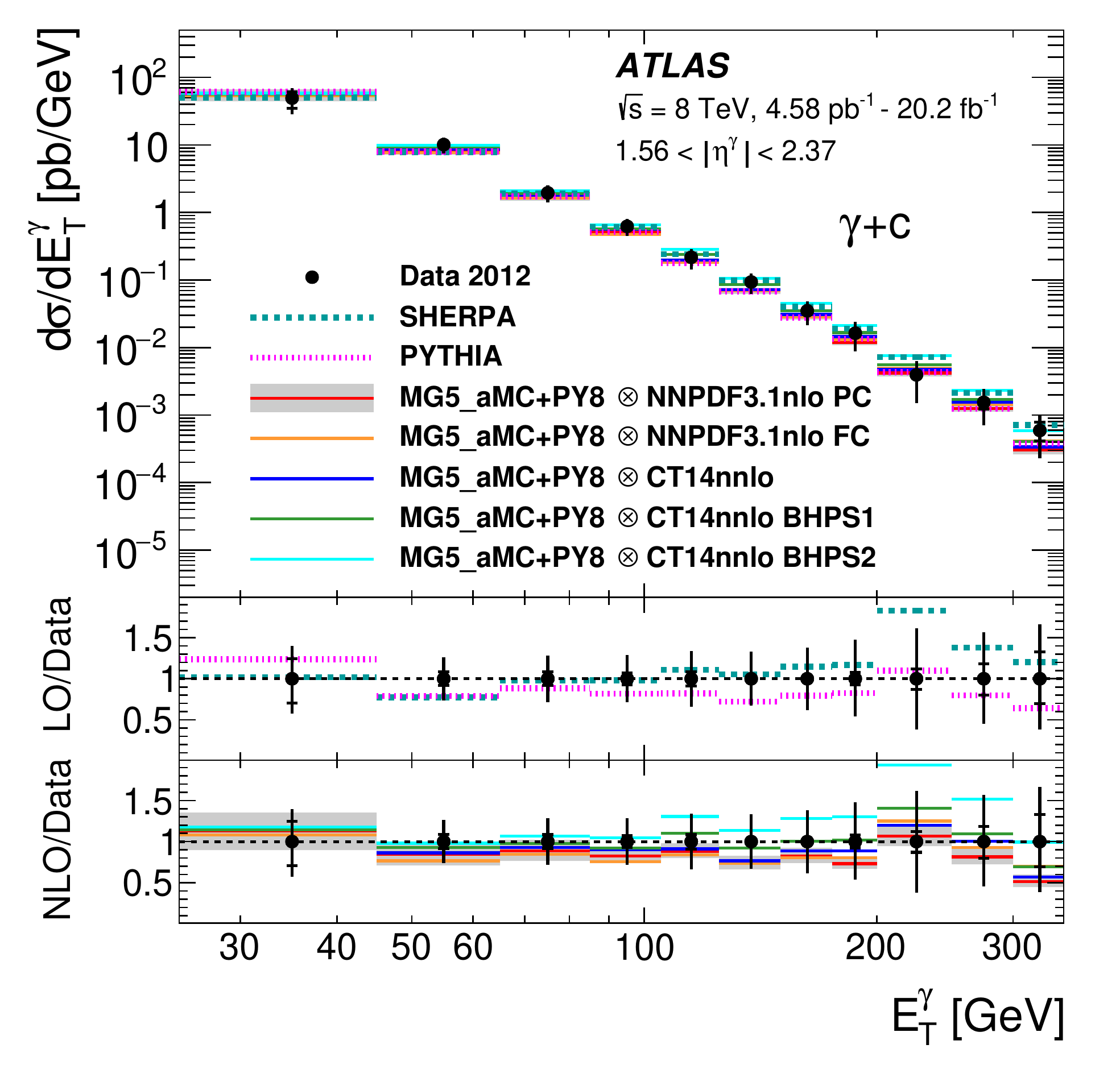}
        \caption{}
    \label{fig:xsecd}
    \end{subfigure}
    \caption{Differential cross sections as a function of the photon transverse energy, \etg, for (a) \yb in the central region, (b) \yb in the forward region, (c) \yc in the central region and (d) \yc in the forward region. The statistical uncertainty is represented as horizontal marks on the error bars of the data points, while the total measurement uncertainty is represented by the complete length of the error bars. The MG5\_aMC+PY8 label in the legend refers to the \MGMCatNLO calculations interfaced to \PYTHIA. The 5F and 4F labels in the legend refer to PDF sets with five quark flavours and four quark flavours respectively. The PC and FC labels in the legend refer to perturbative charm and fitted charm PDF sets respectively. All of the predictions for \yc use PDF sets with five quarks. The theoretical uncertainty in the \MGMCatNLO predictions is displayed for a single PDF set since it is similar for each of the PDF sets. The \SHERPA and \PYTHIA cross sections are not normalized to data and no uncertainties are provided for them.}
    \label{fig:xsec}
\end{figure}

\begin{figure}[htbp]
    \centering
    \begin{subfigure}{0.49\textwidth}
        \includegraphics[width=\textwidth]{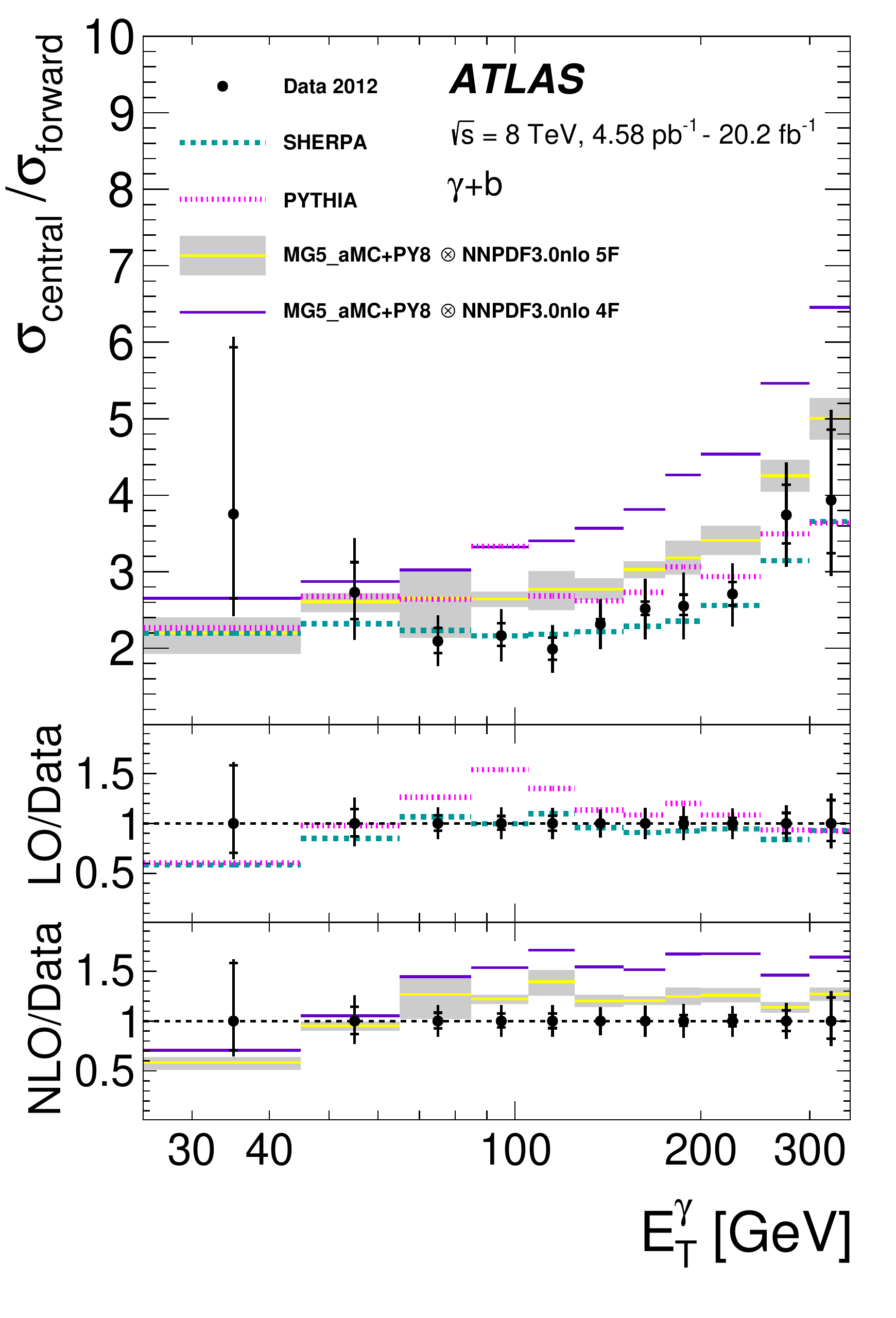}
        \caption{}
    \label{fig:xsecRatioa}
    \end{subfigure}
    \begin{subfigure}{0.49\textwidth}
        \includegraphics[width=\textwidth]{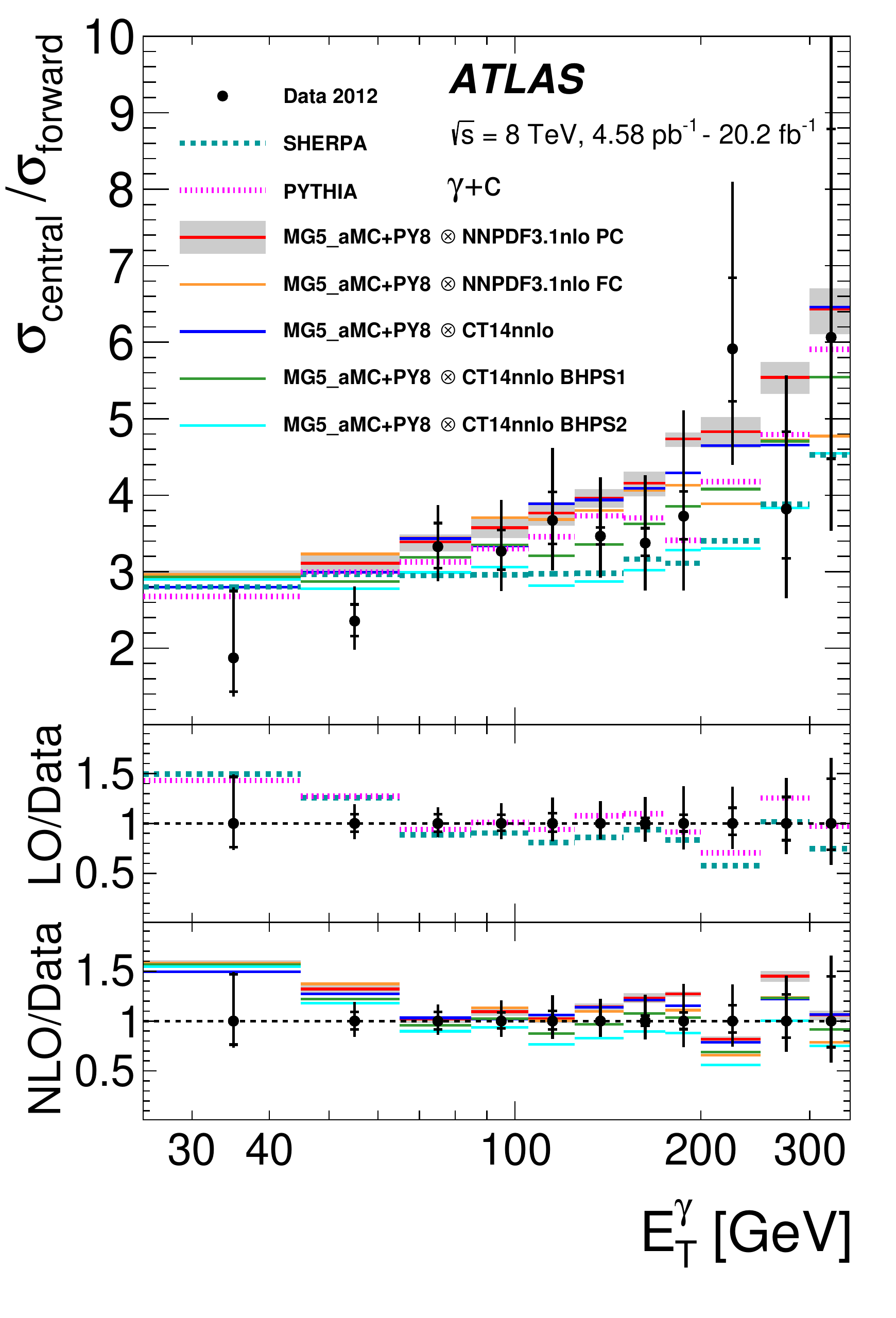}
        \caption{}
    \label{fig:xsecRatiob}
    \end{subfigure}
    \caption{Cross-section ratios of the central region, $|\etag|<1.37$, to the forward region, $1.56<|\etag|<2.37$, as a function of the photon transverse energy, \etg, for (a) \yb and (b) \yc. The statistical uncertainty is represented as horizontal marks on the error bars of the data points, while the total measurement uncertainty is represented by the complete length of the error bars. The MG5\_aMC+PY8 label in the legend refers to the \MGMCatNLO calculations interfaced to \PYTHIA. The 5F and 4F labels in the legend refer to PDF sets with five quark flavours and four quark flavours respectively. The PC and FC labels in the legend refer to perturbative charm and fitted charm PDF sets respectively. All of the predictions for \yc use PDF sets with five quarks. 
The theoretical uncertainty in the \MGMCatNLO predictions is displayed for a single PDF set since it is similar for each of the PDF sets, except for NNPDF3.1nlo FC for which the total uncertainty is similar to that in NNPDF3.1nlo PC at a value of 25~\GeV\ in \etg, but rises steadily relative to it to be a factor of three larger at 350~\GeV. No uncertainties are provided for \SHERPA and \PYTHIA.}
    \label{fig:xsecRatio}
\end{figure}

The values for the measured differential \yb and \yc cross sections and their ratios are given in Table~\ref{tab:diffxsec}. These values are plotted in Figure~\ref{fig:xsec} for the cross sections and in Figure~\ref{fig:xsecRatio} for the ratios, with the relevant theory predictions. In general, considering the LO predictions in pQCD, those from \SHERPA agree well with the measured values and provide a better description of the data than those from \PYTHIA.

\begin{table}[t]
\renewcommand{\arraystretch}{1.7}
\centering
\scriptsize
\caption{Measured values for the \yb and \yc differential cross sections, and their ratios, in the central and forward regions defined respectively as $|\etag|<1.37$ and $1.56<|\etag|<2.37$. They are accompanied by their total measurement uncertainties.}
\label{tab:diffxsec}
\setlength\tabcolsep{2pt}
\begin{tabular}{l r r r r r r r r r r r r}
\toprule
\etg bin [\GeV]&25--45&45--65&65--85&85--105&105--125&125--150&150--175&175--200&200--250&250--300&300--350&350--400\\
\midrule
$\frac{\text{d}\sigma}{\text{d}\etg}$ [pb/\GeV]&$\times10^{0}$&$\times10^{0}$&$\times10^{-1}$&$\times10^{-1}$&$\times10^{-2}$&$\times10^{-2}$&$\times10^{-2}$&$\times10^{-3}$&$\times10^{-3}$&$\times10^{-4}$&$\times10^{-4}$&$\times10^{-4}$\\
\cmidrule(r){1-1}
\cmidrule(l{0.25em}r{0.25em}){2-2}
\cmidrule(l{0.25em}r{0.25em}){3-3}
\cmidrule(l{0.25em}r{0.25em}){4-4}
\cmidrule(l{0.25em}r{0.25em}){5-5}
\cmidrule(l{0.25em}r{0.25em}){6-6}
\cmidrule(l{0.25em}r{0.25em}){7-7}
\cmidrule(l{0.25em}r{0.25em}){8-8}
\cmidrule(l{0.25em}r{0.25em}){9-9}
\cmidrule(l{0.25em}r{0.25em}){10-10}
\cmidrule(l{0.25em}r{0.25em}){11-11}
\cmidrule(l{0.25em}r{0.25em}){12-12}
\cmidrule(l{0.25em}r{0.25em}){13-13}
\yb central&$32^{+\phantom{0}7}_{-\phantom{0}7}$ & $4.6^{+0.7}_{-0.7}$&$11.5^{+\phantom{0}1.5}_{-\phantom{0}1.5}$&$4.0\phantom{0}^{+0.5\phantom{0}}_{-0.5\phantom{0}}$&$14.7^{+\phantom{0}2.2}_{-\phantom{0}2.2}$&$6.8^{+1.0}_{-1.0}$&$3.5\phantom{0}^{+0.5\phantom{0}}_{-0.5\phantom{0}}$&$14.9^{+\phantom{0}2.6}_{-\phantom{0}2.7}$&$6.8^{+1.2}_{-1.2}$&$26\phantom{.0}^{+\phantom{0}5\phantom{.0}}_{-\phantom{0}5\phantom{.0}}$&$8.3^{+\phantom{0}2.1}_{-\phantom{0}2.2}$&$6.0^{+1.5}_{-1.4}$\\
\yb forward&$9^{+\phantom{0}5}_{-\phantom{0}5}$&$1.7^{+0.5}_{-0.4}$&$5.5^{+\phantom{0}0.9}_{-\phantom{0}0.9}$&$1.86^{+0.27}_{-0.27}$&$7.4^{+\phantom{0}1.1}_{-\phantom{0}1.1}$&$2.9^{+0.4}_{-0.4}$&$1.38^{+0.22}_{-0.23}$&$5.9^{+\phantom{0}1.0}_{-\phantom{0}1.0}$&$2.5^{+0.4}_{-0.4}$&$6.9^{+\phantom{0}1.5}_{-\phantom{0}1.4}$&$2.1^{+\phantom{0}0.6}_{-\phantom{0}0.6}$&---\phantom{00}\\
\yc central&$92^{+35}_{-31}$&$24\phantom{.0}^{+5\phantom{.0}}_{-5\phantom{.0}}$&$65\phantom{.0}^{+10\phantom{.0}}_{-10\phantom{.0}}$&$20.4\phantom{0}^{+3.2\phantom{0}}_{-3.1\phantom{0}}$&$80\phantom{.0}^{+13\phantom{.0}}_{-13\phantom{.0}}$&$32\phantom{.0}^{+6\phantom{.0}}_{-6\phantom{.0}}$&$11.8\phantom{0}^{+2.7\phantom{0}}_{-2.6\phantom{0}}$&$61\phantom{.0}^{+15\phantom{.0}}_{-15\phantom{.0}}$&$23\phantom{.0}^{+6\phantom{.0}}_{-6\phantom{.0}}$&$59\phantom{.0}^{+23\phantom{.0}}_{-23\phantom{.0}}$&$36\phantom{.0}^{+13\phantom{.0}}_{-13\phantom{.0}}$&$9\phantom{.0}^{+6\phantom{.0}}_{-6\phantom{.0}}$\\
\yc forward&$49^{+20}_{-21}$&$10.1^{+2.7}_{-2.6}$&$19\phantom{.0}^{+\phantom{0}6\phantom{.0}}_{-\phantom{0}6\phantom{.0}}$&$6.2\phantom{0}^{+1.8\phantom{0}}_{-1.8\phantom{0}}$&$22\phantom{.0}^{+\phantom{0}7\phantom{.0}}_{-\phantom{0}7\phantom{.0}}$&$9.3^{+3.1}_{-3.1}$&$3.5\phantom{0}^{+1.3\phantom{0}}_{-1.3\phantom{0}}$&$16\phantom{.0}^{+\phantom{0}8\phantom{.0}}_{-\phantom{0}8\phantom{.0}}$&$3.9^{+2.4}_{-2.4}$&$15\phantom{.0}^{+\phantom{0}9\phantom{.0}}_{-\phantom{0}8\phantom{.0}}$&$6\phantom{.0}^{+\phantom{0}4\phantom{.0}}_{-\phantom{0}4\phantom{.0}}$&---\phantom{00}\\
\midrule
$\sigma_\text{central}^{\yb}/\sigma_\text{forward}^{\yb}$&$3.8^{+2.3}_{-1.3}$&$2.7^{+0.7}_{-0.6}$&$2.09^{+0.34}_{-0.33}$&$2.17^{+0.35}_{-0.34}$&$1.99^{+0.32}_{-0.31}$&$2.31^{+0.33}_{-0.33}$&$2.5^{+0.4}_{-0.4}$&$2.6^{+0.4}_{-0.4}$&$2.7^{+0.4}_{-0.4}$&$3.7^{+0.7}_{-0.7}$&$3.9^{+1.2}_{-1.0}$&---\phantom{00}\\
$\sigma_\text{central}^{\yc}/\sigma_\text{forward}^{\yc}$&$1.9^{+0.9}_{-0.5}$&$2.4^{+0.5}_{-0.4}$&$3.3\phantom{0}^{+0.5\phantom{0}}_{-0.5\phantom{0}}$&$3.3\phantom{0}^{+0.7\phantom{0}}_{-0.5\phantom{0}}$&$3.7\phantom{0}^{+0.9\phantom{0}}_{-0.7\phantom{0}}$&$3.5\phantom{0}^{+0.8\phantom{0}}_{-0.5\phantom{0}}$&$3.4^{+0.9}_{-0.6}$&$3.7^{+1.4}_{-1.0}$&$5.9^{+2.2}_{-1.5}$&$3.8^{+1.7}_{-1.2}$&$6\phantom{.0}^{+4\phantom{.0}}_{-3\phantom{.0}}$&---\phantom{00}\\
\bottomrule
\end{tabular}
\end{table}

Comparisons of the \yb measurement to NLO+PS predictions from \MGMCatNLO in both the 5F and 4F schemes are shown in Figures~\ref{fig:xseca} and \ref{fig:xsecb}, for the central and forward regions respectively. At low \etg, both the 4F and 5F predictions agree with the data. Above 125~\GeV, however, the 4F predictions underestimate the data. This is consistent with the expectation that 4F is better suited for energies close to the $b$-quark mass. The 5F scheme describes the data better than the 4F scheme at high \etg, with a good description for $\etg<200$~\GeV. However, the 5F scheme underestimates the data at higher \etg values, by up to a factor of two. This is where the gluon-splitting contribution is expected to become more significant relative to the Compton contribution, as the latter depends on the $b$-quark PDF which falls steeply as a function of Bjorken-$x$, and thus \etg. Since the gluon-splitting contribution appears only at tree level in the 5F NLO predictions, this indicates that higher-order calculations would seemingly be needed for a better description of the data in that high \etg region. As shown in Figure~\ref{fig:xsecRatioa}, the 4F and 5F NLO predictions for the cross-section ratios consistently overestimate the data for $\etg>65~\GeV$; the 5F predictions are at the edge of agreement with the measured values within uncertainties. \SHERPA, which generates additional partons in the matrix element and uses a massive 5F scheme, provides a better description of the measured cross sections and cross-section ratios than \MGMCatNLO in either the 5F or 4F scheme.

In comparison to the Compton contribution to the \yb cross section, the Compton contribution to the \yc cross section is larger. This is due to the larger values of both the PDF and absolute electric charge of the $c$-quark, compared to those of the $b$-quark. The gluon-splitting processes, which contribute equally to the \yb and \yc cross sections in the 5F scheme, are thus less important for \yc than for \yb, considering the larger Compton contribution to the former. The gluon-splitting contribution is expected to become important at \etg values around 700~\GeV, beyond the range of this measurement. Comparisons of the \yc measurement to NLO+PS predictions from \MGMCatNLO in the 5F scheme using NNPDF3.1nlo and CT14nnlo are shown in Figures~\ref{fig:xsecc} and \ref{fig:xsecd}, respectively for the central and forward regions. The predictions are found to agree with the data within the uncertainties across the entire \etg range. However, those using the BHPS or the fitted charm PDF sets predict higher cross-section values in the forward region at high \etg, above 105~\GeV, than those using the nominal PDF sets. Correspondingly, the predicted values for the cross-section ratios, shown in Figure~\ref{fig:xsecRatiob}, are smaller for the predictions using the BHPS or the fitted charm sets than for those using the nominal sets. This is the expected behaviour of the intrinsic charm contributions from these PDF sets in the theory predictions. The predictions with the BHPS2 PDF set deviate the most from those using the nominal PDF sets, by about a factor 1.5, while those using the BHPS1 and the fitted charm PDF sets give intermediate values. The precision of the data is comparable to the size of these deviations in the predictions.

Although it is beyond the scope of this Letter, quantitative information about the level of agreement between the data and the theory predictions can be extracted by taking into account the correlations of the measurement uncertainties. Tabulated values of the measurement with full details about their uncertainties and their correlations are provided for this purpose in the Durham HEP database~\cite{hepdata}.

\FloatBarrier

\section{Conclusion}
\label{sec:Conclusion}

Differential cross sections as a function of \etg for isolated prompt photons in association with a \bj or a \cj are measured with the ATLAS detector at the LHC using a dataset of $pp$ collisions at $\rts=8~\TeV$ corresponding to an integrated luminosity of up to 20.2~\ifb. The measured values are compared to LO calculations in pQCD from \SHERPA and \PYTHIA and to NLO calculations in pQCD from \MGMCatNLO interfaced to \PYTHIA. For the \yb final state, the best description of the data is provided by the \SHERPA predictions, which include up to three additional partons and are computed in the massive 5F scheme. The NLO predictions underestimate the data in the highest \etg intervals measured. The 5F scheme of the theoretical calculations provides a better description of the data than the 4F scheme. For the \yc final state, which exhibits larger measurement uncertainties, all the predictions are in agreement with the data. Differences of about the size of the measurement uncertainties are seen between the predictions using PDF sets with intrinsic charm contributions and those without. These measured cross sections provide a test of pQCD calculations with heavy quarks and are sensitive to the $b$- and $c$-quark PDFs.

\section*{Acknowledgements}


We thank CERN for the very successful operation of the LHC, as well as the
support staff from our institutions without whom ATLAS could not be
operated efficiently.

We acknowledge the support of ANPCyT, Argentina; YerPhI, Armenia; ARC, Australia; BMWFW and FWF, Austria; ANAS, Azerbaijan; SSTC, Belarus; CNPq and FAPESP, Brazil; NSERC, NRC and CFI, Canada; CERN; CONICYT, Chile; CAS, MOST and NSFC, China; COLCIENCIAS, Colombia; MSMT CR, MPO CR and VSC CR, Czech Republic; DNRF and DNSRC, Denmark; IN2P3-CNRS, CEA-DRF/IRFU, France; SRNSF, Georgia; BMBF, HGF, and MPG, Germany; GSRT, Greece; RGC, Hong Kong SAR, China; ISF, I-CORE and Benoziyo Center, Israel; INFN, Italy; MEXT and JSPS, Japan; CNRST, Morocco; NWO, Netherlands; RCN, Norway; MNiSW and NCN, Poland; FCT, Portugal; MNE/IFA, Romania; MES of Russia and NRC KI, Russian Federation; JINR; MESTD, Serbia; MSSR, Slovakia; ARRS and MIZ\v{S}, Slovenia; DST/NRF, South Africa; MINECO, Spain; SRC and Wallenberg Foundation, Sweden; SERI, SNSF and Cantons of Bern and Geneva, Switzerland; MOST, Taiwan; TAEK, Turkey; STFC, United Kingdom; DOE and NSF, United States of America. In addition, individual groups and members have received support from BCKDF, the Canada Council, CANARIE, CRC, Compute Canada, FQRNT, and the Ontario Innovation Trust, Canada; EPLANET, ERC, ERDF, FP7, Horizon 2020 and Marie Sk{\l}odowska-Curie Actions, European Union; Investissements d'Avenir Labex and Idex, ANR, R{\'e}gion Auvergne and Fondation Partager le Savoir, France; DFG and AvH Foundation, Germany; Herakleitos, Thales and Aristeia programmes co-financed by EU-ESF and the Greek NSRF; BSF, GIF and Minerva, Israel; BRF, Norway; CERCA Programme Generalitat de Catalunya, Generalitat Valenciana, Spain; the Royal Society and Leverhulme Trust, United Kingdom.

The crucial computing support from all WLCG partners is acknowledged gratefully, in particular from CERN, the ATLAS Tier-1 facilities at TRIUMF (Canada), NDGF (Denmark, Norway, Sweden), CC-IN2P3 (France), KIT/GridKA (Germany), INFN-CNAF (Italy), NL-T1 (Netherlands), PIC (Spain), ASGC (Taiwan), RAL (UK) and BNL (USA), the Tier-2 facilities worldwide and large non-WLCG resource providers. Major contributors of computing resources are listed in Ref.~\cite{ATL-GEN-PUB-2016-002}.

\printbibliography

\clearpage 
\begin{flushleft}
{\Large The ATLAS Collaboration}

\bigskip

M.~Aaboud$^\textrm{\scriptsize 137d}$,
G.~Aad$^\textrm{\scriptsize 88}$,
B.~Abbott$^\textrm{\scriptsize 115}$,
O.~Abdinov$^\textrm{\scriptsize 12}$$^{,*}$,
B.~Abeloos$^\textrm{\scriptsize 119}$,
S.H.~Abidi$^\textrm{\scriptsize 161}$,
O.S.~AbouZeid$^\textrm{\scriptsize 139}$,
N.L.~Abraham$^\textrm{\scriptsize 151}$,
H.~Abramowicz$^\textrm{\scriptsize 155}$,
H.~Abreu$^\textrm{\scriptsize 154}$
% ...
\bigskip
\\
% ...
$^{12}$ Institute of Physics, Azerbaijan Academy of Sciences, Baku, Azerbaijan\\
% ...
$^{88}$ CPPM, Aix-Marseille Universit{\'e} and CNRS/IN2P3, Marseille, France\\
% ...
$^{115}$ Homer L. Dodge Department of Physics and Astronomy, University of Oklahoma, Norman OK, United States of America\\
% ...
$^{119}$ LAL, Univ. Paris-Sud, CNRS/IN2P3, Universit{\'e} Paris-Saclay, Orsay, France\\
% ...
$^{137}$ $^{(a)}$ Facult{\'e} des Sciences Ain Chock, R{\'e}seau Universitaire de Physique des Hautes Energies - Universit{\'e} Hassan II, Casablanca; $^{(b)}$ Centre National de l'Energie des Sciences Techniques Nucleaires, Rabat; $^{(c)}$ Facult{\'e} des Sciences Semlalia, Universit{\'e} Cadi Ayyad, LPHEA-Marrakech; $^{(d)}$ Facult{\'e} des Sciences, Universit{\'e} Mohamed Premier and LPTPM, Oujda; $^{(e)}$ Facult{\'e} des sciences, Universit{\'e} Mohammed V, Rabat, Morocco\\
% ...
$^{139}$ Santa Cruz Institute for Particle Physics, University of California Santa Cruz, Santa Cruz CA, United States of America\\
% ...
$^{151}$ Department of Physics and Astronomy, University of Sussex, Brighton, United Kingdom\\
% ...
$^{154}$ Department of Physics, Technion: Israel Institute of Technology, Haifa, Israel\\
$^{155}$ Raymond and Beverly Sackler School of Physics and Astronomy, Tel Aviv University, Tel Aviv, Israel\\
% ...
$^{161}$ Department of Physics, University of Toronto, Toronto ON, Canada\\
% ...
$^{a}$ Also at Department of Physics, King's College London, London, United Kingdom\\
$^{b}$ Also at Institute of Physics, Azerbaijan Academy of Sciences, Baku, Azerbaijan\\
$^{c}$ Also at Novosibirsk State University, Novosibirsk, Russia\\
$^{d}$ Also at TRIUMF, Vancouver BC, Canada\\
$^{e}$ Also at Department of Physics {\&} Astronomy, University of Louisville, Louisville, KY, United States of America\\
% ...
$^{*}$ Deceased
\end{flushleft}


\end{document}